# Spin labeling - electron paramagnetic resonance spectroscopy study of temperature and salt dependent serum albumin stability

Sergei P. Rozhkov, Andrey S. Goryunov

Institute of Biology, Karelian Research Center RAS, Petrozavodsk, Russia

Electron paramagnetic resonance (EPR) signal parameters of TEMPO-dichlorotriazine and TEMPO-maleimide spin labels attached to human serum albumin (HSA) molecules have been monitored at various temperatures, salts and sucrose concentrations. The method for measuring the correlation time $\tau^{eff}$ of the protein rotational dynamics using viscosity isotherms was utilized. $\tau^{eff}$ is proportional to the effective protein volume $V_M^{eff}$ dependent on protein structure dynamic stability in a certain temperature range which can thus be determined. Sharp temperature changes in the effective protein volume and their dependence on the concentration of NaCl, $(NH_4)_2SO_4$, polyethylene glycol, and heavy water are interpreted as inter-conversions of stable low- and high-temperature protein conformers with the participation of intermediates. Few variants of the qualitative phase diagrams of water-salt dispersion of HSA were presented on the basis of the results obtained. The regions have been marked, where thermo- and salt-induced liquid-liquid phase transitions can occur, as well as intermediates and supramolecular structures of the protein can form in their course.

**Introduction**

A concept of thermodynamic stability of the structure of protein molecule based presently on the ideas of protein folding is widely used to describe the properties of globular proteins in solutions and dispersions in a wide temperature range [1-3]. As follows from the bell-shaped form of the protein stability curve, destabilization of the native state occurs at negative values of the difference in the free energy $\Delta G^D_N = G^D - G^N$ between the native N and denatured D protein states below and above the two 1st order phase transition temperatures of thermal and cold denaturation, where $\Delta G^D_N = 0$. The shape of the stabilization curve changes when the structure of protein or water in solution alters in the presence of osmolytes or adsorbents [2,3], which biological activity can thus be studied .

The thermodynamic stability of protein dispersion should also be considered in relation to diffusion processes [4,5]. Phase stability of the dispersion is violated (concentration inhomogeneities cease to "dissolve") at $\delta^2 G = 0$, where $\delta^2 G = \det(\partial^2 G/\partial m_i/\partial m_j)$, and $m_i$ и $m_j$ are the molar concentrations of the solution components: protein and osmolyte, respectively. This determinant allows describing the boundary of the system stability to phase separation (spinodal) [6]. The condition for the existence of a critical point of the liquid-liquid type phase transition (LLPT) was determined using the phase diagram in the temperature - salt concentration coordinate plane. LLPTs is usually observed in the immediate vicinity of the solubility curve of the protein native form, when the molecules are able to aggregate, crystallize, and polymerize [7,8]. At the same time the conformational variability of protein molecules, as well as the possibility of protein structural states, which are characterized by a partial structural disorder [9] and can be considered as intermediate between the native and denatured states, were not yet taken into account. The simultaneous presence of native and denatured molecules can also lead to a decrease in the thermodynamic stability of the protein solution as a whole [10] and promote phase separation.

The separation of protein into fractions can arise due to the presence of molecules with an uneven distribution of electrolyte ions adsorbed in their structure [11]. An important factor in the formation of partially disordered regions in a protein globule can be conformational changes induced by interaction with metal ions [12], contributing to the manifestation of protein structural flexibility. Thus, LLPT can be associated with the composition of the protein dispersion, which simultaneously contains a stable native form and conformational intermediates [12-14] in a temperature range near the 1$^{st}$ order PT of both thermal and cold denaturation.

Despite numerous studies of water-salt dispersions of serum albumin (SA), no phase diagram (PD) of the protein in the temperature - salt concentration coordinate plane or stability curves were presented so far. This is probably due to the extreme structural heterogeneity of the SA [15]. In the presence of ligands and salts, SA dispersion is assumed to be an equilibrium mixture of macromolecules with varying degrees of liganding in different macrostates. Binding of ions and ligands changes the charge equilibrium in the protein and, as a result, the interdomain interactions and the conformation of the protein globule. In addition SA cold denaturation is strongly biased towards negative temperatures [2], which does not allow plotting of a stability curve.

An approach based on the spin labeling in combination with electron paramagnetic resonance (EPR) spectroscopy has been used in this study with the aim to assess changes in the protein stability in the region of $\Delta G^D_N$ maximum values under the effect of a number of osmolytes. To this end we propose a variant of the method using a variable sucrose concentration. Sucrose affects the viscosity of the protein dispersion and at the same time maintains an ordered native state of the protein [16,17]. This leads to a decrease in its structural flexibility and an increase in its effective volume $V_M$ in comparison with the disordered state. This is reflected in the broadening of the EPR spectra and an increase in the protein rotational correlation time $\tau_M$. Therefore, the increase in dispersion viscosity, plotted for a specific segment of the stabilization curve at each individual temperature, allows a consistent assessment of the corresponding increase in effective protein volume. The closer to the top of the stabilization curve, the smaller the increment of the stabilization curve, which is inversely proportional to the effective volume $V_M$, coupled with the segmental mobility (flexibility) of the protein. As the stability of the native state decreases (temperature shifts towards both cold and heat denaturation), the protein flexibility increases, and the effective volume $V_M$ decreases (the correlation time of the protein rotational dynamics $\tau_M$ decreases). This is registered by the ESR spin labeling method. In this study we also analyze the section of the SA stability curve of in the region of its $\Delta G^D_N$ maximum values in the range of physiological temperatures in the presence of a number of osmolytes, without using the construction of the stability curve in the whole temperature range.

EPR spin labeling is one of the methods that allow studying in parallel both the conformational state of protein molecules, intermolecular interactions and the formation of complexes (associates) of protein molecules in solution [18,19]. Therefore, we are also aimed at detecting changes in the structural-phase state of SA solutions, caused by the effect of salts and temperature on the conformation of proteins, in order to clarify a more general question about the features of the phase states of SA solutions basing on the current understanding of the phase properties of protein systems. The issue of the position of the stability regions of protein conformers and associates on the phase diagram, as well as phase transitions between these states, including LLPT, are discussed in particular.

## Materials and Methods

### Materials

Human serum albumin (HSA) (fraction V, fatty acid containing) lyophilized powder was purchased from Sigma-Aldrich Company and used without further purification. TEMPO-maleimide spin label purchased from Sigma-Aldrich and TEMPO-dichlorotriazine (2,2,6,6 - tetramethyl - N - 1 - oxylpiperidine - 4 - aminodichlorotriazine) spin label synthesized by the method described by Zhdanov [20] were used for spin labeling of BSA molecules in aqueous dispersion. Sucrose, polyethylene glycol MM 40,000 (PEG-40), NaCl, $(NH_4)_2SO_4$ were of analytical grade; and heavy water (deuterium oxide, $D_2O$) 99.8 atom % D. Ultrapure distilled, degassed and deionized Millipore water (18.2 MΩ cm resistivity) was used in all preparations of phosphate buffer solutions (pH 7.3) containing 0.15 M NaCl and HSA solutions prepared on this basis. Freshly prepared solutions of SA were used in all sets of experiments. Solutions of spin-labeled HSA (HSA-SL) for EPR spectroscopy were obtained by diluting of 1 mM initial protein solution in 0.01 M PBS, containing 0.15 M NaCl (pH 7.3) to the required concentration. Sample volumes were 100-300 μL. Solutions with different contents of heavy water were prepared by adding a concentrated protein solution to buffer solutions with an appropriate concentration of $D_2O$. pH of all samples was carefully monitored using a calibrated pH microelectrode and a pH meter. Measurements in solutions containing $D_2O$ were carried out at pH equal to the measured pH to ensure that the electrostatic state of the macromolecules remains unchanged. The viscosity of sucrose solutions was determined using an Abbe refractometer (IRF-22, RF) and a corresponding nomogram. The viscosity of combined solutions of sucrose with PEG-40 was determined using an Ostwald viscometer.

**Methods**

*EPR spin label instrumentation*

EPR spectra of HSA-SL were recorded on an EPR X-band Bruker EMX 6/1 radiospectrometer with a thermostated cavity cell (± 0.2 °C). We used modulation amplitude of 1 G and microwave power of 12.6 mW to avoid saturation and signal distortion. Modification of HSA with spin labels were described in detail earlier [21-23]. The resulting spectrum of a spin label for its slow rotation is the combination of two well-resolved external extrema caused by hyperfine interaction and a central component caused by the superposition of various components of the hyperfine structure. A spin label can have both several binding sites on a protein (the case of TEMPO-dichlortriazine spin label [19]), or one binding site (proxyl-maleimide [24]). The spin label ESR spectrum contains several components in both cases. The two-component spectrum of the TEMPO-maleimide spin label was found to be due to the equilibrium of two local conformational states of HSA structure [25]. The equilibrium constant K between the states with different microviscosity and mobility of the spin label is determined by conformational fluctuations of the cavity in the structure of the protein, where the spin label is localized. The changes in the effective entropy $\Delta S^{eff}$, enthalpy $\Delta H^{eff}$ and free energy of the cavity state were calculated and shown to depend on the concentration of electrolyte and protein [22,23]. A structural change characteristic of LLPT in the physiological temperature range was also found.

The correlation time calculated for the spin label diffusion depends on the selected model of its dynamics and reflects the diffusion of both the spin label itself with a frequency $\nu_1 = \tau_R^{-1}$, and a protein globule with a frequency $\nu_2 = \tau_M^{-1}$. This correlation time $\tau^{eff} \sim \nu_1^{-1} + \nu_2^{-1}$ is effective and can be calculated using the well-known formula [18,19]:

$$\tau^{eff} = a\,(1-S)^b\,32/A_z, \qquad (1)$$

where $S = A_z^*/A_z$, $A_z$ is a half of the field distance between the outer extrema of the spectrum corresponding to the condition of extremely slow rotation of the spin label (77K), and $A_z^*$ is the same value for the spectrum of the spin label under study. Parameters $a$ and $b$ are determined by the spin label rotation model and the residual width of the individual spectral line, and within the framework of the model used (isotropic or weakly anisotropic jump-like diffusion) their values are $a = 25.5\,10^{-14}$ T / s and $b = -0.615$. This diffusion model was chosen because the rotation patterns of the protein and the label are similar. The "jumps" of the label itself are caused by conformational changes in its microenvironment, and the "jumps" of the macromolecule are caused by the long-range repulsive and short-range attractive potentials of protein-protein interaction. The potential barrier of the interaction can only be overcome by a "jump" of the macromolecule resulting in both translational diffusion and rotational reorientation:

$$\nu = (\tau_{R+M}^{eff})^{-1} = \tau_R^{-1} + \tau_M^{-1} \qquad (2)$$

Within the framework of an isotropic model of the spin label motion, the effective correlation time $\tau^{eff} = \tau_{R+M}$ can be divided into components $\tau_R$ and $\tau_M$, using the dependence of the resulting rotational frequency of the spin label on the ratio $T/\eta$, where $\eta$ is the viscosity of the solvent [26]. Taking into account the Stokes-Einstein law for rotational diffusion of a macromolecule, equation (2) can be written as follows:

$$\tau_{R+M}^{-1} = \tau_R^{-1} + 3kT/4\pi V_M \eta, \quad (3)$$

$V_M$ is the effective protein volume. At a constant temperature, the $\tau_R$ value is determined by the effective microviscosity of the protein matrix ($\eta_2^{eff}/T$), coupled with surface tension [27], and $\tau_M$ is determined by the viscosity of the solvent ($\eta/T$). The microviscosity of the water-protein matrix is two orders of magnitude higher than the bulk solvent viscosity, so the latter can hardly affect the dynamics of the spin label in an aqueous protein solution. Thus, the expression for the slope of the dependence $\tau_{R+M}^{-1}$ on $T/\eta$ is as follows:

$$\Delta(\tau_{R+M}^{-1})/\Delta(T/\eta) = 3k/(4\pi \tau_M^{eff}) \quad (4)$$

An approximation of the linear section of the dependence $\tau_{R+M}^{-1}$ от $T/\eta$ to infinite viscosity ($T/\eta \rightarrow 0$) gives $\tau_{R+M}^{-1} = \tau_R^{-1}$.

At the same time, the slope of the extrapolation lines depends on the value of the effective protein volume $V_M$, which is associated with the segmental protein mobility. The latter, in turn, is determined by the course of the protein stability $\Delta G^D_N$ curve versus temperature. The closer the maximum of the stability curve (stability of the native state), the greater the value of $V_M$ [28]. As the stability of the native state decreases (the temperature shifts to the region of both cold and thermal denaturation), the volume $V_M \sim \tau_M$ decreases (mobility increases). Stabilization of the protein structure under the effect of sucrose can be represented as an increase in the effective temperature $T/\eta$, and the accompanying change in $\Delta G^D_N$ can be considered as an increment of the stabilization energy $|\Delta G^D_N(T+\Delta(T/\eta)) - \Delta G^D_N(T)|/\Delta(T/\eta)$. $V_M \sim \tau_M^{eff}$ takes the maximum value in the region of the maximum of the stabilization energy curve and this can be registered by the above mentioned EPR spin labeling method using viscosity isotherms.

The changes in the average distances R between paramagnetic centers in vitrified (77 K) protein dispersions can be registered from the dipole-dipole broadening of the spectral lines using the parameter $(R)^{-1} \sim d_1/d$, where $d_1/d$ is the ratio of the total intensity of the extreme spectrum components to the intensity of the central components [19]. To this end the samples were kept preliminarily at a temperature corresponding to a certain section of the stability curve and then immersed immediately into the liquid nitrogen. Each phase state of a dispersion in a corresponding temperature range of phase stability is assumed to be characterized by its certain average distribution of spin label dipoles on the protein molecules and the corresponding average distance between them.

The research was carried out using the equipment of the Core Facility, Karelian Research Centre RAS.

**Results**

The structural-dynamic state of albumin molecules is characterized by a number of reversible conformational transitions in the region of non-physiological pH values [29], which are also successfully recorded using ESR spin labeling [24]. An increase in temperature to 55° C causes a gradual and reversible change in the protein secondary structure. The most significant changes caused by the unfolding of one of the protein domains occur at 42-43 °C [30,31]. A sharp onset of denaturation processes with the formation of starting aggregates takes place in the region 58–65 °C [32]. Cold denaturation occurs in the region of negative °C temperatures. However, the LLPT with the upper critical solution temperature (UCST) at positive °C temperatures was observed in the presence of polyethylene glycols (PEG) [33]. On the other hand, the reentrant LLPT with the LCST in the physiological temperature range has been recorded in the presence of multivalent salts.

SA in solution can be represented by the native conformer A in the temperature range 12-20 °C, and by two native conformers A and B in the temperature range 22-50 °C, the proportion of conformer B increasing with temperature [34]. Spontaneous transformation of A into B occurs at 58 °C and higher. These data on A and B conformers also correlate with the presence of two maximum correlation times of spin-labeled albumin in the physiological temperature range in the presence of $D_2O$ [21]. The stability of the protein is known to increase under the presence of $D_2O$, which indicates the important role of hydration in stabilization [35]. In addition, $D_2O$ contributes to the stabilization of the monomeric form of the protein [36]. Conformers A and B can cause corresponding stabilization curves instead of one common curve, what is best manifested in the presence of $D_2O$.

Fig. 1 shows the temperature dependences of the effective volume $V_M^{eff} \sim \tau_M^{eff}$ of HSA-SL molecules in the phosphate buffer solution containing 0.15 M NaCl, as well as in the presence of $D_2O$ and PEG-40. The values are normalized to the maximum $V_M^{eff}$ of HSA-SL corresponding to the maximum of the stabilization curve. $\tau_M^{eff}$ was calculated using formula (4) based on the parameters of the EPR spectra obtained for the TEMPO-dichlorotriazine spin label bound to HSA. All the dependences are non-monotonous, but the dependence for NaCl shows a pronounced maximum at around 33 °C, while there are two temperature maxima at around 15 and 33° C for $D_2O$ (Fig. 1 (2)). Changes in $V_M^{eff}$ in the range from 15 to 33 °C can be considered as occurring in the interval between the stabilization peaks of A and B of protein conformers.

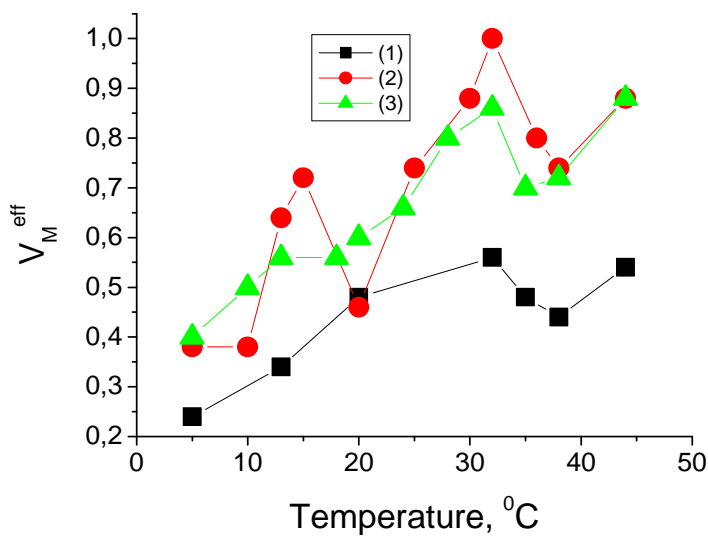

**Fig. 1** Temperature dependences of the normalized effective rotational volume $V_M^{eff} \sim \tau_M^{эф}$ of HSA-SL molecules spin labeled with a TEMPO-dichlorotriazine in 0.01 M phosphate buffer solution pH 7.3 containing: 0.15 M NaCl and 2% PEG-40 (1); 0.15 M NaCl and 5% $D_2O$(2); 0.15 M NaCl (3). HSA concentration is 13 mg/ml

Matsarskaia et al. [37] reported an LLPT with LCST in the SA dispersion in the presence of polyvalent salts, with the transition temperature decreasing with increasing salt concentration. This indicates a decrease in the thermodynamic stability of the dispersion, which leads to phase separation. The dispersion of the protein B-conformer in this case can be considered a stable phase. The clusters of the remaining protein A-conformers that have lost solubility and stability in this temperature and salt concentration range can be considered as the metastable phase within the stable phase. The effect of the solvent ionic strength on the protein conformers has been studied by increasing the salt concentration in order to reveal the relationship of these transformations with the state of the HSA dispersion. NaCl and $(NH_4)_2SO_4$ have been compared as the salts that differ in their ability to alter stability and phase state of HSA. The "salting out" effect of the latter salt is known to be higher.

The temperature dependences of the normalized $V_M^{eff}$ value at various contents of NaCl and $(NH_4)_2SO_4$ (Fig. 2) are increasing dependences for the average salt concentration range. The difference with those shown in Fig. 1 is that the position of the transition shifts to lower temperatures. The low-temperature maximum becomes higher with a further increase in the salt concentration, and $V_M^{eff}$ values are lower at higher temperatures. This may be due to a further destabilization of the protein A-conformer and a shift of LLPT to lower temperatures.

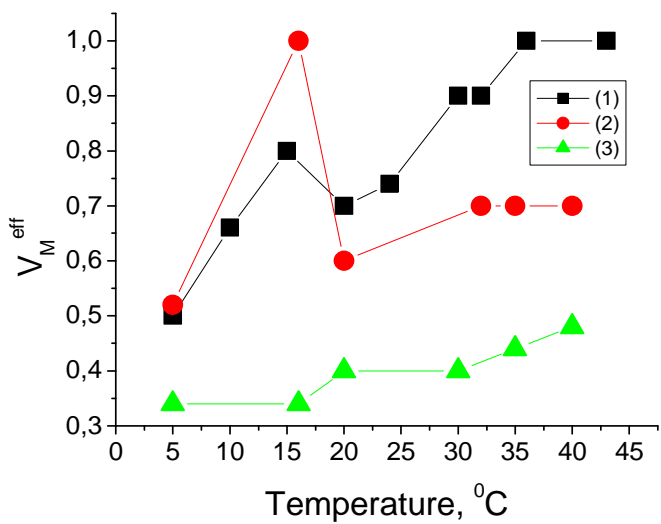

**Fig. 2** Temperature dependences of the normalized value of the normalised effective rotational volume $V_M^{eff} \sim \tau_M^{эф}$ of HSA-SM molecules modified with the TEMPO-dichlorotriazine spin label in 0.01 M phosphate buffer solution pH 7.3 containing: 0.75 M NaCl (1); 0.15 M NaCl and 0.6 M $(NH_4)_2SO_4$ (2); 0.15 M NaCl and 1 M $(NH_4)_2SO_4$ (3). HSA concentration is 13 mg/ml

$V_M^{eff}$ values are lower throughout the temperature range at salting-out concentrations (Fig. 2 (3)). $V_M^{eff}$ significantly decreases ($\tau_M^{eff}$ = 12 ns) at 5 °C and 1 M $(NH_4)_2SO_4$ and the phase separation of the solution is visually observed as the formation of the protein gel. The same happens in the presence of PEG-40 (Fig. 1 (1)). These results may indicate that the thermodynamic stability of the HSA-SL dispersion is significantly reduced over the entire temperature range up to gel formation.

The $\tau_M^{eff}$ value of 12 ns at 5 °C was recorded previously in our study of solutions of spin-labeled antibodies (IgG-SL) in the presence of PEG and dextran polymers, and the formation of the gel was also visually observed [unpublished results]. IgG belongs to the protein class with UCST [38]. This means that L-L phase separation in its dispersion can be caused by a decrease in temperature, and the critical point shifts towards higher temperatures with increasing PEG concentration. Thus, a low-temperature transition with UCST is observed for both HSA-SL and IgG-SL in the presence of PEG.

The data obtained indicate that the protein stability decreases over the entire temperature range in the presence of salting-out concentrations, and it is most likely that the LCST and UCST are getting closer in this case. A decrease in the correlation time can be explained by the coexistence of protein molecules in the initial native state and more labile intermediate conformers under the conditions of phase separation. They are grouped into clusters of a separate phase during the LLPT with both LCST and UCST. In general, the stability of the native state decreases, and the effective volume of a protein molecule $V_M^{eff} \sim \tau_M^{eff}$ is determined by the self-diffusion of individual structural protein domains, rather than the macromolecule as a whole.

On the other hand, the very shape of the EPR spectra of the spin label under gel conditions still indicates a significant immobilization of the label due to the surface-inactive properties of the salts. This leads to an increase in the surface tension and stabilization of the intrinsic protein dynamics in the vicinity of the spin-label localization sites [27] and is evidenced by the increased correlation times $\tau_R$ (Eq. 3).

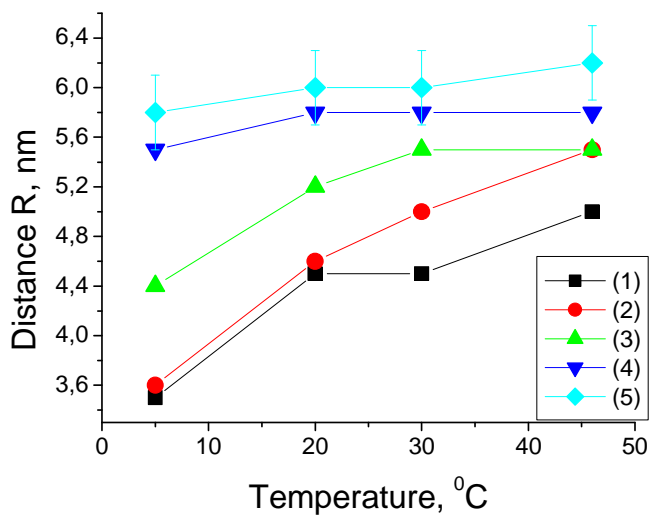

**Fig. 3** Temperature dependences of the average distance R between the paramagnetic sites of TEMPO-maleimide spin labels attached to HSA in 0.001 M phosphate buffer solution (pH 7.3) containing different NaCl concentrations: 0.001 M (1); 0.01 M (2); 0.15 M (3); 0.5 M (4); 1.2 M (5). HSA concentration is 50 mg/ml. Temperature 77 K

Fig. 3 shows the temperature dependences of the average distance R between the paramagnetic sites of the TEMPO-maleimide spin label attached to HSA for different NaCl concentrations in the protein dispersion. The data were obtained by measuring the dipole-dipole interaction between spin labels under conditions of complete freezing of the diffusion of protein macromolecules at 77 K [19]. The interaction is determined by the distance between the spin labels $d_1/d \sim R^{-1}$, and hence between the macromolecules carrying them.

It follows from Fig. 3 that the average distance between spin labels is 1–2 nm larger in high salt protein solutions than in low salt solutions. It is also seen that phase rearrangements starts in the range of average salt concentrations at high temperatures. The effects of temperature and salt concentration can add up and affect the hydration barrier that stabilizes the protein A-conformer. Therefore, the structure of the solution determined by the stable A-conformer of HSA at higher temperatures is likely to change first. It is also known that the fraction of albumin oligomers can be up to 20% at room temperatures [29]. Probably these oligomers are formed mostly by A-conformers and fall apart gradually on rising temperature and salt concentration. This can explain an increase in the average distance of the dipole-dipole interaction. Such a destruction of oligomers would first promote an increase in the concentration of monomers in the A-conformation (an increase in $V_M^{eff}$ at low temperatures in Fig. 2) at moderate salt concentrations. This can be also followed by the transformation of A-conformers into both B-conformers and high-temperature intermediates I*, which can further participate in high-temperature LLPT with LCST. Previously, we reported the changes in the effective entropy $\Delta S^{eff}$, enthalpy $\Delta H^{eff}$ and free energy associated with the A and B conformational transitions of HAS. These depended on the molar concentrations of the protein $m_2$ and electrolyte $m_3$, $m_2/m_3 = $ const [6]. This is typical of LLPT [37].

**Discussion**

The protein stabilization curve $\Delta G = \Delta G^D - \Delta G^N$ and its derivative $\partial \Delta G / \partial T = -\Delta S^D_N$ for two native protein A and B conformers, stable in their respective temperature ranges, can be represented schematically as follows (Fig. 4).

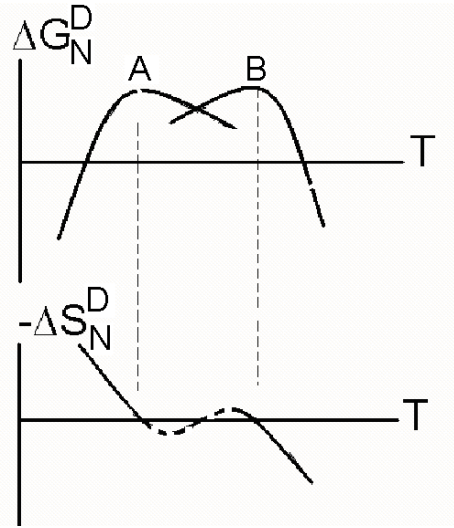

**Fig. 4** Protein stabilization curve $\Delta G = \Delta G^D - \Delta G^N$ and its derivative $\partial \Delta G / \partial T = -\Delta S^D_N$ for two native protein A and B conformers differing in the temperature of stabilization maxima

It is noteworthy that two points $\Delta S^D_N = 0$ of the lower graph (Fig. 4) correspond to the maxima of the stabilization curve. At these points, the native and denatured states are energetically indistinguishable, judging by the zero difference in entropy. However, the difference in the volume of the native, denatured and intermediate molecules causes a difference in their chemical potentials due to the Laplace pressure. The larger the volume, the lower the chemical potential and vice versa. Thus, the native state is preferable and dominant. In this case, "hotter" intermediate molecules with a higher chemical potential should tend to combine into a cluster microphase in order to compensate for the difference in chemical potentials with their native conformers. This can lead to LLPT. An important factor in the formation of partially folded intermediates can be metal- or ion-induced conformational changes in protein [12]. We have previously analyzed the issue of the energy favorability of protein clustering and microphase formation [22]. We assumed that polar or charged groups of one protein molecule located in the region of non-polar groups of another protein can affect the state of water in the region of the non-polar groups and thereby increase the entropy. The greater the compensation effect, the fewer protein molecules are needed to form a cluster. Proteins with an increased exposure of non-polar groups to bulk water (presumably intermediates) lose entropy and are preferable for the formation of clusters. Intermediates that differ less from native protein molecules, for example due to thermally induced structural changes, are also under less driving force for L-L phase separation. This is equivalent to a critical phase transition and the formation of a macroscopically homogeneous solution in a certain temperature range.

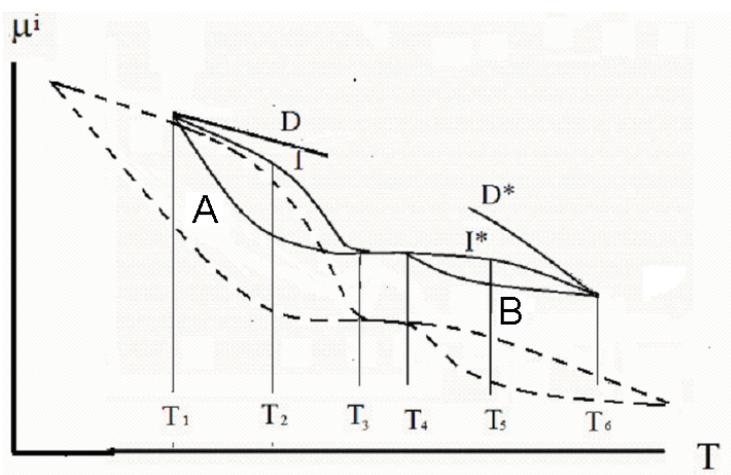

**Fig. 5** Schematic drawing of hypothetic temperature dependences of chemical potentials $\mu^i$ of protein conformers (i = D, D *, A, B, I, I *). $T_1$ and $T_6$ are the points of N↔D and N*↔D* $1^{st}$ order phase transitions; $T_2$ and $T_5$ are optimal temperatures for the clusters formation; $T_3$ and $T_4$ are the points of the critical phase transitions between the phases of native proteins and clusters of intermediates with UCST and LCST, respectively. $T_3$-$T_4$ interval is the supercritical macroscopically single-phase region. Addition of salts shifts $T_1$ and $T_6$ towards lower and higher temperatures (dotted lines)

Let us consider the possible behavior of the chemical potentials $\mu^i$ of low-temperature native A, denatured D, and intermediate I protein states, as well as their corresponding high-temperature analogs—B, D*, I* states—in the temperature range between cold and heat denaturation (Fig. 5). Understanding that D and D * states, although close, are not identical [39], we also assume that A conformer gradually transforms to B conformer.

The chemical potential $\mu$ of a substance decreases with rising temperature at constant pressure, since $(\partial\mu/\partial T)_p = -S$, entropy S is positive [40]. Fig. 5 shows that states A and B are the most stable states of the protein ($\mu^A$ and $\mu^B$ are the lowest) in the temperature range between $T_1$ and $T_6$, while states D and D * are the least stable states ($\mu^D$ and $\mu^{D*}$ are the highest). Accordingly, the concentration of protein in a more stable state increases, and in states D * and D it becomes negligible. Then the chemical potentials of the protein intermediates I and I *, as well as their concentrations, have intermediate values. At temperatures $T_1$ and $T_6$ (1$^{st}$ order phase transition), the equality of chemical potentials ensures the equilibrium of all states of the protein. The difference between the chemical potentials of native proteins and intermediates is most significant at around temperatures $T_2$ and $T_5$, where the destabilization of a single-phase solution and LLPT can occur. $\Delta S^I_N$ equals zero in the temperature range $T_3$ -$T_4$ and the difference between all molecules disappears. These are critical PTs. There is a region of continuous phase transitions in the range between $T_3$ and $T_4$ [41], where gradual interconversions of various conformers take place.

Fig. 6 shows an extended phase diagram of protein dispersion in temperature-composition coordinate plane based on the data presented in Figs. 1 - 3 and on the scheme of Fig. 5. The composition is described here by the ratio $m_2/m_3$ of the molar concentrations of protein and salt, respectively [6].

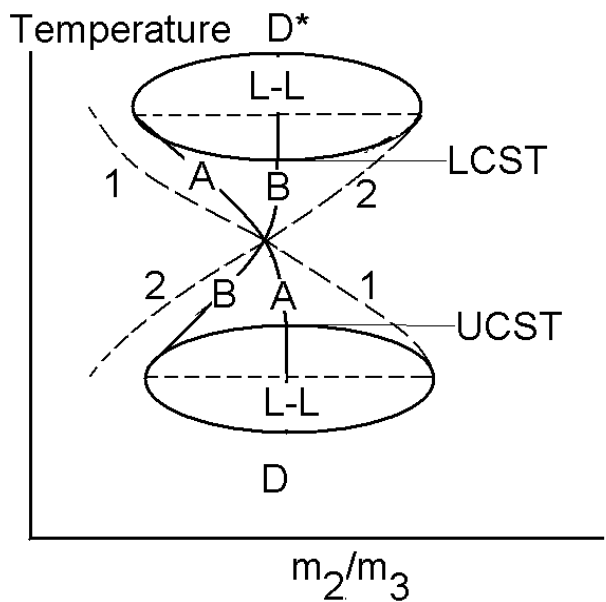

**Fig. 6** A schematic drawing of the hypothetic phase diagram of protein dispersion in the temperature - composition coordinate plane. The composition is described by the ratio $m_2/m_3$ of the molar concentrations of protein and salt, respectively. A and B are the solubility curves of the native A and B protein conformers. Dashed curves denote the solubility of protein oligomers (1) and curvilinear fibrils (2). Horizontal dotted lines are the bimodal regions with UCST and LCST

The diagram pictures two LLPTs [33], the presence of two interconverting conformational states A and B of the native protein as well as of a large fraction of protein oligomers [29] (curve 1) and ordered aggregates of amyloid-like fibrils [41,42] (curve 2). Both aggregates and fibrils can arise in the interval between UCST and LCST as a possible result of destabilization of low-temperature protein clusters and high-temperature linear fibrils formed during LLPTs.

Solubility lines A and B may intersect at intermediate temperatures, since µA exceeds µB, and conformer A increasingly converts to B form with rising temperature. However, some molecules can still retain the conformation A, albeit with reduced solubility. They form intermediate I*. $\mu^B$ exceeds $\mu^A$ at low temperatures and conformer A prevails in the protein solution. In this case, the remains of the conformer B form intermediate I. The region of conformers forming a solution is to the left of these lines, and the region of condensation states is to the right.

**Conclusions**

The results of an EPR spin labeling study of the structural-dynamic properties of human serum albumin in aqueous dispersions of various compositions have been presented. The viscosity isotherms of the correlation time of the spin label dynamics were used to estimate the variations in the effective rotational volume of the protein molecule and the temperature ranges of the maximum stability of the native state of the protein. The dependences of these maxima on the concentration of sodium chloride, ammonium sulfate, polyethylene glycol, and heavy water, as well as the temperature dependences of the effective protein volume proportional to the protein correlation time, can be explained by interconversions of low- and high-temperature protein conformers with the possible formation of intermediate products. Changes in the average distance between HSA molecules, which we observed in an aqueous protein dispersion under the effect of an increase in the concentrations of sodium chloride and ammonium sulfate, suggest the stabilization of the low-temperature conformer, including through the salt-induced degradation of protein oligomers.

Further destabilization of the A-conformer is accompanied by its transition to the B-state and the formation of high-temperature intermediates. This can lead to the formation of a dense phase, represented by high-temperature intermediates, during the reentrant LLPT with LCST, while the main phase is represented by B-conformers.

The regions of thermal and salt-induced LLPTs were determined using the phase diagram (PD) of the states of an aqueous protein dispersion, proposed on the basis of the results obtained. The regions indicate conditions for protein conformers, clusters, oligomers and other aggregates formed in the course of LLPT.

Thus, using the EPR spin label method, not only conformational changes determining the temperature ranges of protein stability, but also the phase properties of protein dispersions associated with these changes were characterized. This allows us to expect that the method and the proposed approach will be effective in studying the mutual influence of proteins and nanoparticles of different chemical nature with the aim of developing a generalized PD of globular proteins taking into account the conformational, including denatured, state of the protein.

**Acknowledgment**